\documentclass[aps,pra,twocolumn, showpacs,floatfix,superscriptaddress]{revtex4-1}
\usepackage{graphicx,amsmath,amssymb}
\usepackage{bm}
\usepackage{hyperref}
\usepackage{color}
\usepackage{graphicx}
\usepackage{dcolumn}
\usepackage{braket}
\usepackage{soul} % added by Philippe
\usepackage{upgreek}
\usepackage{xcolor}
\usepackage[utf8]{inputenc} % Acentos: cedilha, til, crase, circunflexo
\usepackage{ulem} % tachar texto % comando \sout{}

\begin{document}

\title{Temporal dynamics of L\'evy flights of photons in a three-dimensional geometry for hot vapor}

\author{R.\,V.\,M.~de Almeida Filho}
\affiliation{Departamento de F\'isica, Universidade Federal de Pernambuco, 50670-901, Recife, Pernambuco, Brazil}

\author{J.\,C.~de Aquino Carvalho}
\affiliation{Departamento de F\'isica, Universidade Federal de Pernambuco, 50670-901, Recife, Pernambuco, Brazil}

\author{T.~Passerat de Silans}
\affiliation{Departamento de F\'isica, CCEN, Universidade Federal da Para\'iba, Caixa Postal 5008, 58051-900, Jo\~ao Pessoa, Para\'iba, Brazil}

\author{M.\,H.\,G.~de Miranda}
\affiliation{Departamento de F\'isica, Universidade Federal de Pernambuco, 50670-901, Recife, Pernambuco, Brazil}

\author{M.\,O.~Ara\'ujo}
%\email{michelle.oaraujo@ufpe.br}
\affiliation{Departamento de F\'isica, Universidade Federal de Pernambuco, 50670-901, Recife, Pernambuco, Brazil}
 
\date{\today}

\begin{abstract}
Multiple scattering of light by resonant vapor is characterized by Lévy-type superdiffusion with a step size distribution $P(x) \propto 1/x^{1+\alpha}$, with $0 < \alpha < 2$. The L\'evy parameter $\alpha$ was measured from $P(x)$, steady fluorescence, frequency-dependent fluorescence and time-resolved transmission, %\st{all of them in the forward direction} 
where the latter was observed for quenched disordered media, not yet for annealed disordered media. %Here we report first measurements of this quantity from time-resolved backward fluorescence, i.e., photons that are backward diffused from light pulses exciting a hot rubidium vapor. We show experimentally that $\alpha$ can be extracted from this diffuse reflection, and the results are consistent with time-resolved transmission (i.e., photons that are forward diffused) and steady frequency-dependent forward fluorescence.
Here we report first measurements of this quantity from time-resolved fluorescence for photons that are forward diffused from light pulses exciting a hot rubidium vapor, a medium characterized by annealed disorder, where correlations between the photon steps are absent. Also, the vapor is contained in a three-dimensional geometry, a turned cylinder with respect of the incident photon direction. We show experimentally that $\alpha$ can be extracted from this setup, and the results are consistent with those given by the steady frequency-dependent forward fluorescence. Theoretical simulations are consistent with these results. Also, our simulations show the validity of the one-dimensional theories for our actual geometry and other three-dimensional ones, as thin and thick disks, and a measurement of the number of steps that the photon performed before leaving the medium.%\st{although we measure $\alpha=1$ for both transmission and reflection, the backward photons have a non-negligible amount of single scattering events even for high density, contrary to the forward photons where multiple scattering dominates.}
\end{abstract}

%\keywords{Suggested keywords}%Use showkeys class option if keyword %display desired
\maketitle

%\tableofcontents

%---------------------------------------------
\section{Introduction}

Random walk of particles have applications in a large variety of systems, with the Brownian motion of particles the most known example. This motion is characterized by a random walk with successive jumps of size $x$, performed with finite step-size variance within a Gaussian distribution. However, in the last decades, random walks with diverging step-size variance were extensively studied, known as L\'evy flights or L\'evy walks. For this kind of random walk, the probability distribution for the step size $x$ is given by $P(x)\propto 1/x^{\alpha+1}$, where $0<\alpha < 2$ is the L\'evy parameter. L\'evy flights or walks are encountered in several kind of systems, ranging from animal foraging~\cite{Viswanathan1996, Edwards2007}, phonon transport in nanowires~\cite{Li2022} and isolation of individuals between populations~\cite{Smith2023}.

Light propagation in turbid media is often described by standard random walk and diffusion equations, like in %cloudy atmospheres~\red{[citation needed]}, 
biological tissues~\cite{Chernomordik2002} and cold atoms~\cite{Labeyrie2003}. However, in astrophysics~\cite{Ivanov2024}, photonic materials dubbed L\'evy glasses~\cite{Barthelemy2008} and hot atomic vapors~\cite{Mercadier2009}, %\red{Versão de Marcio} 
L\'evy flights %in atomic vapors 
can provide the correct statistical framework to describe superdiffusive transport of photons. In astrophysics and hot vapors, the L\'evy statistics arises from multiple resonant scattering events and cannot be fully described by normal Brownian diffusion. In this process, photons are absorbed and re-emitted by the atoms with a different frequency which depends on the atomic velocity and collisions. Because of Doppler and eventual collisional broadening, 
%the re-emitted photon frequencies follow a Doppler or Lorentzian distribution (in fact, a Voigt profile)}, respectively, making its frequency to be non-correlated with the input frequency (complete  frequency redistribution) and allowing some photons in the far spectral wings to travel long distances before being reabsorbed~\cite{Holstein1947, Holstein1951, AlvesPereira2007}. 
the re-emitted photons have their frequncies non-correlated with the input frequency (i.e., complete frequency redistribution), which allows some photons to appear in the far spectral wings and they travel long distances before being reabsorbed~\cite{Holstein1947, Holstein1951, AlvesPereira2007}. 
These long and rare flights produce a power-law step size distribution with $\alpha=1$ ($\alpha=0.5$) for non-collisional (collisional) broadening, defining the experimental signature of L\'evy flights and superdiffusion. Modelling photon propagation in alkaline vapor as a L\'evy flight gives an accurate physical description of radiation trapping~\cite{Chevrollier2012} and enables precise experimental studies of light transport and L\'evy parameters under tunable conditions \citep{Macedo2021}. %It is also essential for understanding and optimizing applications such as high-resolution laser spectroscopy~\red{[citar outro paper]}.

Signatures of L\'evy flights in atomic vapors were observed from the direct measurement of the step length distribution~\cite{Mercadier2009}, the radial diffuse transmission~\cite{Baudouin2014, Araujo2021} and laser-frequency-dependent diffuse transmission~\cite{Macedo2021, Lopez2023}. Time-resolved transmission was observed in the 1980s in sodium vapor for broad excitation~\cite{Colbert1990} and data were analysed with transport radiative transfer equation~\cite{Holstein1947, Holstein1951}. On the other hand, experiments with ultrashort laser pulses were performed in L\'evy glasses~\cite{Savo2014}, where the walk dimension was measured. This medium is structured, which fractal dimension, quenched disorder, and introduces correlations between successive steps, which makes the formalism of L\'evy walk more appropriate. In all these works, it was considered a slab geometry for the medium and the output photons were collected at the surface opposite to the laser input, i.e., photons in the forward direction or diffuse transmission. Theories on L\'evy flights for different geometries and multidimensional media are still scarce in the literature~\cite{Zaburdaev2015}, although recent works discuss L\'evy flights in disks and spheres~\cite{Klinger2025} and fractals in two and three-dimensional L\'evy-like random walks~\cite{Chalas2026}. %It is known that for cold atoms, most of the photons \red{ver: Considering a slab geometry, most of the photons leave the medium in the backward direction, according with the called Ohm law [citation needed]}. 
%Up to our knowledge, time-resolved transmission was only observed in the forward direction.

In this work, we investigate the forward time-resolved scattered photons, i.e., the diffuse transmission, for photons in a hot atomic Rb vapor. Contrary to L\'evy glasses, hot vapors have annealed disorder, i.e., absence of correlations between the random steps and the random walk of the photon is modelled by a L\'evy flight instead of a L\'evy walk. We use a non-usual geometry, more precisely, a turned cylinder cell, i.e., a cylinder whose main axis is perpendicular to the incident laser. We extract the L\'evy parameter and clearly distinguish between single and multiple scattering, where the number of scattered photons is, respectively, small and large. For comparison, we also present results for the total diffuse transmission by scanning the incident laser detuning. Monte Carlo simulations confirm and support our experimental results. Since our system has a bounded geometry, our results provide access to the distribution of the number of scattering events that the photon performed before leaving the medium. This is the first-passage time (or exit time) for the photons, a quantity not yet observed for vapors. In addition, we show that our results are valid for other three-dimensional geometries, like a thin slab and a thick slab, which validate analytical equations derived for one-dimensional L\'evy flight.

%Considering a slab geometry, most of the photons leave the medium in the backward direction, according with the called Ohm law~\red{citation needed}. Up to our knowledge, time-resolved transmission was only observed for diffuse transmission, i.e., the fluorescence emitted in the forward direction. 
%In this work, we report an investigation of the backward time-resolved transmission, i.e., the diffuse reflection, of photons an hot atomic Rb vapor. We extract the L\'evy parameter $\alpha$ and we clearly distinguish between single scattering and multiple scattering, where the photon scattering number is, respectively, small and large. For comparison, we also present results for the forward time-resolved transmission (diffuse transmission) and for the total diffusing transmission by scanning the incident laser direction. We perform an (...) This study provides a direct observation of signatures of L\'evy for the backward emitted fluorescence where the signal-to-noise ratio is high, and (...).

%---------------------------------------------
\section{Theory}

We consider an atomic vapor at a temperature $T$ and density $n=n(T)$. After a photon has been absorbed by an atom and emitted by spontaneous emission in a random direction, it travels a distance $x$ whose probability distribution is given by $P(x)$. The photon is emitted with a new frequency which depends on the incident frequency and the atom velocity. Assuming that the photon performs $N_{sc}$ steps before leaving the vapor, for large $N_{sc}$ we have complete frequency redistribution~\cite{Pereira2004} and $P(x)$ is given by

\begin{equation}
P(x)=\int_{-\infty}^{\infty} k^2(\Delta) e^{-k(\Delta)x}\,d\Delta \xrightarrow{x\rightarrow\infty} \dfrac{1}{x^{\alpha+1}} \, \text{,}
\label{Px}
\end{equation}
\\*
where $\Delta=\omega-\omega_0$ is the photon detuning, with $\omega$ as the photon frequency and $\omega_0$ as the atomic resonance frequency, $k(\Delta)=n\sigma_0 f(\Delta)$ is the absorption coefficient, $n$ is the atom density and $\sigma_0$ is the resonant cross section. $f(\Delta)$ is the line shape, a Voigt profile given by %\red{Chev2013}

\begin{equation}
f(\Delta)\propto \dfrac{1}{\sqrt{\pi}u} \int_{-\infty}^{\infty} \dfrac{e^{-v^2/u^2}}{1+4(\Delta-kv)^2/\Gamma^2} \, dv \,.
\label{eq_fV}
\end{equation}

In Eq.~\ref{eq_fV}, $v$ is the component of the atom velocity parallel to the photon, $u=\sqrt{2k_B T/M}$  is the most probable velocity, $M$ is the atomic mass, $T$ is the vapor temperature, $k_B$ is the Boltzmann constant and $\Gamma$ is the atomic natural linewidth. Eq.~\ref{eq_fV} takes into account the atom natural lineshape and the Doppler broadening. Approximating the Voigt profile core by a Doppler profile and from Eqs.~\ref{Px} and~\ref{eq_fV}, the parameter $\alpha$, sometimes called L\'evy index or L\'evy exponent, is equal to $\alpha=1$. 
For collisional broadening, Eq.~\ref{eq_fV} is replaced by a Lorentzian function with width $\Gamma_\mathrm{col}\gg \Gamma$ and leads to Eq.~\ref{Px} with $\alpha=0.5$. %\red{Acho importante  pelo menos citar o caso colisional em uma frase, porque aqui \'e uma revisao teorica e a mencao \'e por completeza.}

If the vapor cell has a slab geometry whose walls are at the positions $z=0$ and $z=L$ along an $z$ axis, and if photons enter at $z_0 \gtrsim 0$ ($0 < z_0 \ll L$), the temporal photon intensity profile at the forward exit surface ($z=L$) is called diffuse transmission. For large $t$, it is given by an exponential decay~\cite{Baudouin2014, Savo2014, Klinger2023}

\begin{equation}
T_{\mathrm{diff}}(t)\sim e^{-t/\tau}\, ,
\label{Tdiff_temp}
\end{equation}
\\*
where~\cite{Savo2014}

\begin{equation}
\tau \propto L^d\,,
\label{tau_Ld}
\end{equation}
\\*
and $d$ is called the walk dimension with $d=\alpha$~\cite{Klinger2023}. The exponential decay in Eq.~\ref{Tdiff_temp} was also reported in theories with transport equation~\cite{Holstein1947,Holstein1951}, where the time decay rates were given as $\tau\propto b_0 \sqrt{\ln b_0} \sim b_0$ for Doppler broadening, % and $\tau\propto \sqrt{n}=b^{0.5}$ in the $R_{III}$ scenario, 
where $b_0=n\sigma_0 L$ is the resonant optical depth of the vapor.

%On the other hand, the temporal intensity profile at the backward exit surface ($z=0$) is called diffuse reflection, and for large $t$ is given by~\cite{Klinger2023}

%\begin{equation}
%R_{\mathrm{diff}}(t)\sim e^{-t/\tau}\,,
%\label{Rdiff_temp}
%\end{equation}
%\\*
%i.e., the same exponential dependence as $T_\mathrm{diff}(t)$. This
%Equation~\ref{tau_Ld}} suggests that it is possible to observe L\'evy signatures from the forward time dependent} transmission, and this is the aim of the present work.
The aim of the present work is to provide the observation of L\'evy signatures from the forward time dependent transmission for a hot vapor. %Also, %as discussed in subsection~\ref{subsec_results}, a good signal-to-noise ratio is expected for photon collection in the backward direction, due to the so-called Ohm's law for photons~\cite{VanRossum1999, Guerin2017}, i.e., $T_\mathrm{diff}\rightarrow 0$ for large $L$ for atoms in a slab geometry, which implies in increasing $R_\mathrm{diff}$ for large $L$. 
We replace $L$ by the vapor density $n$ in Eq.~\ref{tau_Ld}, since we keep $L$ constant and vary $n$. %since for large $L$ the majority of the photons are backward scattered. This is  as already discussed in previous works~

% For cold atoms in a spherical geometry, as the sample radius increases, the majority of the photons are backward scattered and few are transmitted (see Fig. 4 of~\cite{Labeyrie2004}), and we checked that this is also the case for hot vapors. This means that a good signal-to-noise ratio is expected for photon collection in the backward direction.%, contrary to the forward direction at high vapor density.

Equations~\ref{Tdiff_temp} and~\ref{tau_Ld} are derived for L\'evy flights in one dimension~\cite{Klinger2023} for an infinite slab geometry~\cite{Baudouin2014}, although for infinite vertical cylinder ref.~\cite{Holstein1951} also finds theoretically the scaling $\tau\propto b_0 \sqrt{\ln b_0} \sim b_0$. As it will be discussed in the next section, our geometry is a finite turned long cylinder, i.e., a finite three-dimensional system. In the next sections we report successfully the validity of these results in our experimental data and simulations.

%Finally, it is valid to mention the Sparre-Andersen theorem: for large $t$, we have the universal scaling  $R_{\mathrm{diff}}(t)\sim 1/t^{3/2}$~\cite{Klinger2023}, i.e., exits from the forward direction are improbable. As it will be discussed in section~\ref{sec_sim}, our experimental data and simulations show this behavior: for increasing density, the time photons spend in the vapor increases and the backward fluorescence increases compared with the forward fluorescence.}

%$t\ll n^\alpha$, $T_{\mathrm{diff}}(t) \sim 1/t^{1/2}$ and $R_{\mathrm{diff}}(t)\sim 1/t^{3/2}$~\cite{Klinger2023}, i.e., a universal scaling of the temporal [ver o limite de slab infinitamente espesso]}

%---------------------------------------------
\section{Experiment}

\subsection{Setup}
\label{subsec_setup}

\begin{figure}[h]
\centering
\includegraphics[scale=0.8]{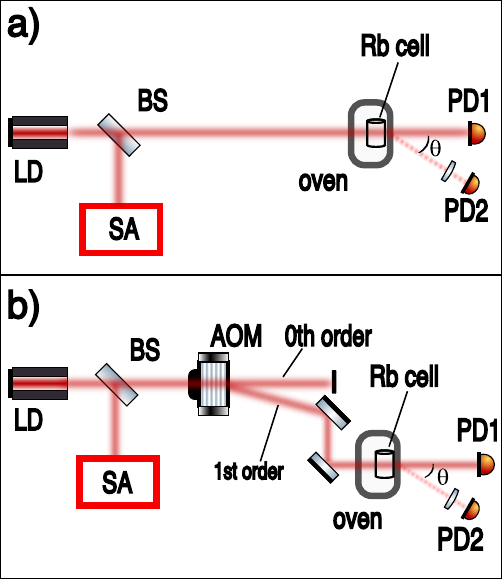}
\caption{(a) Scheme of the experimental setup. A saturated-absorption spectroscopy (SA) is used for frequency calibration. The laser goes to a Rb cell in an oven. The photodetectors PD1 and PD2 measures, respectively, the coherent transmission and the diffuse transmission, for scanning frequency. (b) The laser diode is frequency-stabilized and an AOM produces pulses. Its first-order beam drives the vapor in order to measure the temporal diffuse transmission.}
\label{fig1}
\end{figure}

Our experimental setup consists of a heated rubidium vapor cell, illuminated by a low-power homemade diode laser, resonant with the D2 line of rubidium ($\lambda=780.24$ nm) and frequency stabilized by a saturated absorption setup [see Fig. \ref{fig1}(a)]. The laser beam is sent to a monomode PM optical fiber (not shown). %, where a $TEM_{00}$ Gaussian mode is obtained. 
At the cell entrance, the laser beam is linearly polarized and has a diameter of $2$ mm and %a power of 
$50$ $\mu$W power. The photodetectors PD1 and PD2 are detectors from ThorLabs, model APD120A/M, and collect the laser transmission (PD1; coherent transmission) and the emitted fluorescence (PD2) in the forward direction (diffuse transmission). %; Figs.~\ref{fig1}(a) and (b)] or in the backward direction [Fig.~\ref{fig1}(c)]. 
This fluorescence is made of photons that performed a random walk in the vapor, and are collected by PD2 by means of a $30$ mm focal lens, placed at a distance of approximately $10$ cm from the oven aperture and at an angle of $\theta=30^\circ$ from the laser direction. This angle is relatively large in order to minimize the scattered light by cell walls from coherent transmission. The cell is cylindrical where its main axis is placed perpendicularly to the laser beam direction. The cell has a length of $9$ cm and a radius $1.25$ cm. This cell setup was necessary because when directing the laser along the cell length, we observed a small diffuse transmission, i.e., practically no photons passing through the end. The cell is placed in a homemade oven that heats the entire unit. The oven has a circular aperture of diameter $2$ cm (not shown in Fig.~\ref{fig1}), from where the coherent transmission and the fluorescence exit.

The laser coherent transmission as a function of the incident laser detuning $\Delta$ is theoretically adjusted in order to determine the vapor density $n$. For this, we adjust the coherent transmission (measured by photodetector PD1) with the Beer-Lambert law, $T_\mathrm{coh}(\Delta) \propto e^{-n\sigma(\Delta)L}$, where the density $n$ is kept as the free parameter and the Voigt cross section is equal to $\sigma(\Delta)=\int_{-\infty}^{\infty} \sigma_L(\Delta-kv)W(v)\,dv$, where $W(v)=e^{-v^2/u^2}/\sqrt{\pi u^2}$ is the normalized Maxwell velocity distribution for the component $v$ of the atom velocity, parallel to the laser direction, and $\sigma_L(\Delta)$ is the Lorentzian cross section and takes into account the whole hyperfine structure of $^{85}$Rb and $^{87}$Rb. It is given by~\cite{Mercadier2013}

\begin{eqnarray}
\sigma_L(\Delta)&=&\dfrac{\lambda^2}{\pi} \left[ \sum_{F=2}^3 \sum_{F'=1}^4 \dfrac{2F+1}{\sum_{F_1=2}^3 (2F_1+1)} \right. \nonumber \\
&\times& \dfrac{S_{FF'}}{1+4(\Delta-\omega_{FF'})^2/\Gamma^2} + \nonumber \\
&+& \sum_{F=1}^2 \sum_{F'=0}^3 \dfrac{2F+1}{\sum_{F_1=1}^2 (2F_1+1)} \nonumber \\
&\times& \left. \dfrac{S_{FF'}}{1+4(\Delta-\omega_{FF'})^2/\Gamma^2} \right] \, ,
\label{eq_rb}
\end{eqnarray}
\\*
where $S_{FF'}$ are transition factors calculated from the Clebsch-Gordan and the Wigner 3-$j$ coefficients~\cite{Steck2001} and $\omega_{FF'}$ is the transition frequency from the level $F$ to $F'$. In the right-hand side of Eq.~\ref{eq_rb}, the first (second) term within the brackets represents the sum over the hyperfine levels of $^{85}$Rb ($^{87}$Rb). 
%The laser coherent transmission as a function of the incident laser detuning $\Delta$ is theoretically adjusted by using the equations of ref.~\cite{Mercadier2013} in order to determine the vapor density $n$. For this, we adjust the coherent transmission (measured by photodetector PD1) with the Beer-Lambert law, $T_\mathrm{coh}(\Delta) \propto e^{-n\sigma(\Delta)L}$, where the density $n$ is kept as the free parameter and the cross section $\sigma(\Delta)$ takes into account the whole hyperfine structure of $^{85}$Rb and $^{87}$Rb.
Temperatures range from $20^\circ$C to $145^\circ$C, which correspond to densities from $4.6\times 10^{15}$ to $4.2\times 10^{19}$ atoms/m$^3$, respectively. Errorbars for the density were calculated for a set of few measurements with same temperature, and correspond to approximately $15\%$ of its mean value.

%A densidade tambem foi uma media. No fim a barra de erro da densidade eh 15\%, e do tau ficou 20\%.

In order to measure the temporal dynamics of the diffuse reflection and transmission, the laser has its frequency stabilized in the transition $F=2\rightarrow F'=1,2,3$ of $^{87}$Rb. % by a lock-in amplifier from ????, and 
The vapor is driven by square pulses of  duration $16$ $\mu$s, generated by the $+1$ order of an acousto-optical modulator (AOM) from Isomet, model 232A-1 [see Fig.~\ref{fig1}(b)]. The $+1$ order corresponds to $80$ MHz above the transition $F=2\rightarrow F'=3$ of $^{87}$Rb. Note that due to the large Doppler broadening of the vapor, typically $\sim 500$ MHz, the exact value of the photon input frequency is not important, since it will be resonant with the vapor energy level or line inside this width. The AOM is placed at the focus of a telescope made of two lenses of focal distances $10$ mm, in order to improve the pulse extinction time. %Two detection schemes were employed: diffuse transmission [Fig.~\ref{fig1}(b)] and diffuse reflection [Fig.~\ref{fig1}(c)]. 
Each measured fluorescence for a fixed density is averaged for 2,000 cycles, corresponding to 2$-$3 hours of data acquisition. %The errorbars were (...) \red{Falar também das barras de erro} [cada tau eh uma media de $\approx 400$ ciclos, e o desvio padrao tambem. A densidade tambem foi uma media. No fim a barra de erro da densidade eh 15\%, e do tau ficou 20\%.]

%\subsection{Experimental results}

%\subsection{Diffuse transmission versus frequency}

% Fig2a: T_diff vs Delta
% Fig2b: T_diff vs n
\begin{figure}[h]
\centering
\includegraphics[scale=1]{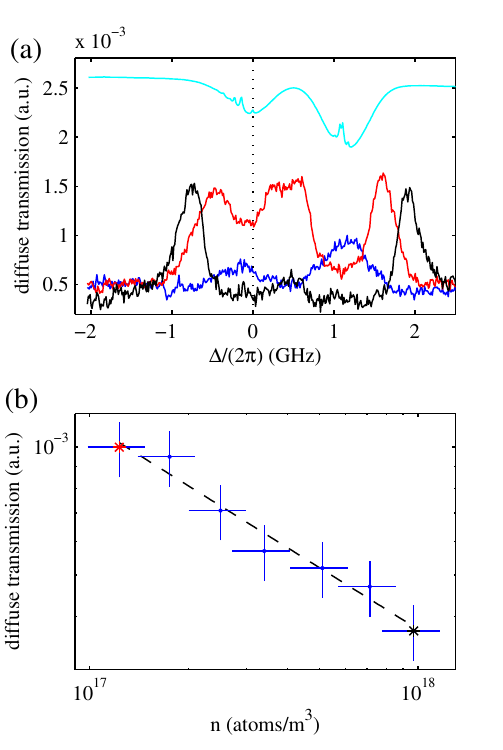}
\caption{(a) Diffuse transmission $T_\mathrm{diff}(\Delta)$ as a function of the laser detuning for three different temperatures: $30^\circ \text{C}$ (blue; $n=9.5\times 10^{15}$ atoms/m$^3$), $70^\circ \text{C}$ (red; $n=1.8\times 10^{17}$ atoms/m$^3$) and $100^\circ \text{C}$ (black; $n=1.6\times 10^{18}$ atoms/m$^3$) The top cyan curve is the Rb saturated absorption (S.A.). All data were obtained with the setup of Fig.~\ref{fig1}(a). (b) Measured diffuse transmission $T_\mathrm{diff}(0)$ as a function of sample density $n$ (blue dots). The corresponding temperatures are, in $^\circ$C: $70$, $75$, $80$, $85$, $90$, $95$ and $100$. The two asterisks represent $T_\mathrm{diff}(0)$ extracted from (a). The black dashed line is a fit of the experimental data with Eq.~\ref{Tdiff_vs_n}.}
\label{fig2}
\end{figure}

% Fig3: T_diff vs n
%\begin{figure}[h]
%\centering
%%\includegraphics[scale=0.55]{figure3.pdf}
%\includegraphics[scale=1]{figure3_eu.pdf}
%\caption{Measured diffuse transmission amplitude as a function of sample density. Dashed line is a fit to experimental data.}
%\label{fig3}
%\end{figure}

\subsection{Calibration}

To validate our experimental setup, we use the setup of Fig.~\ref{fig1}(a) to measure $\alpha$ using the method of ref.~\cite{Macedo2021}, done for a Cs vapor, which consists in scanning the incident laser frequency $\omega$ in order to measure the frequency-dependent diffuse transmission $T_\mathrm{diff}(\Delta)$. Then we collect $T_\mathrm{diff}(0)$, % i.e., at $\Delta=0$, 
where $\Delta=0$ corresponds to the transition $F=2 \rightarrow F'=3$ of $^{87}$Rb, and we plot it as a function of the vapor density $n$.

The diffuse transmission as a function of the detuning is shown in Fig.~\ref{fig2}(a) for different temperatures when the laser is scanned around the transition $F=2 \rightarrow F'=1,2,3$ of $^{87}$Rb D2 line. Note that $T_\mathrm{diff}(0)$ is not equal to the minimum of $T_\mathrm{diff}(\Delta)$ because the reference transition is close to the transition $F=3 \rightarrow F'=2,3,4$ of $^{85}$Rb, so there is an overlap of wings of both Rb isotopes at $\Delta=0$. %We checked this effect in simulations \red{To do?}.  The Rb saturated absorption profile is also shown for reference.
Fig.~\ref{fig2}(b) shows $T_\mathrm{diff}(0)$ plotted as a function of the vapor density $n$, collected from $T=70^\circ$C ($n=1.8\times 10^{17}$ atoms/m$^3$) to %\st{$T=110^\circ$C ($n=3.3 \times 10^{18}$ atoms/m$^3$)} 
$T=100^\circ$C ($n=1.6 \times 10^{18}$ atoms/m$^3$). Between $30^\circ$C and $70^\circ$C, $T_\mathrm{diff}(0)$ \textit{versus} $n$ increases, and above $110^\circ$C it is constant [not shown in Fig. 2(b)], due to low photon scattering number and detector noise level, respectively.

It was shown~\cite{Baudouin2014, Araujo2021, Macedo2021} that $T_\mathrm{diff}(0)$ scales with $n$ according to

\begin{equation}
T_\mathrm{diff}(0) = \frac{C}{n^{\alpha/2}}\,,
\label{Tdiff_vs_n}
\end{equation} 
\\*
where $C$ is a positive constant. We adjust our data with Eq.~\ref{Tdiff_vs_n} by keeping $C$ and $\alpha$ as free parameters. This gives
$\alpha=1.0 \pm 0.1$, %fit de Ricardo
%$\alpha=0.9 \pm 0.3$, %meu fit
i.e., consistent with $\alpha=1$. Errorbars for $T_\mathrm{diff}(0)$ were evaluated as the same way as those for $n$, i.e., by averaging a set of few measurements with same temperature and are approximately $20\%$ of its mean value for a fixed $n$.

To investigate the possibility of collisions in our data, we estimate the atom-atom collisional rate as

\begin{equation}
\Gamma_\mathrm{col}=\beta n\,,
\label{gamma_c}
\end{equation}
\\*
where $\beta=9\times 10^{-8}$ Hz$\cdot$cm$^{3}$~\cite{Weller2011}. For the range of densities in Fig.~\ref{fig2}(b), this gives $\Gamma_\mathrm{col}$ between %\red{valores corrigidos} 
$15$ kHz (for $n\approx 2 \times 10^{17}$ atoms/m$^3$) and $300$ kHz (for $n\approx 3 \times 10^{18}$ atoms/$m^3$). Simulations for $T_\mathrm{diff}$ for several densities were done from $T=50^\circ$C to $T=130^\circ$C, which gives $\Gamma_\mathrm{col}$ between $4$ kHz and $1$ MHz (see details about the simulation methods in section~\ref{sec_sim}). Also, we performed simulations by superestimating  Eq.~\ref{gamma_c} by a factor of five and ten. In the simulations, $T_\mathrm{diff}$ is made of all photons scattered in the forward direction for $\theta\in[0^\circ,90^\circ]$ relative to the incidence direction (note that in the experiment $\theta$ is fixed at $\theta=30^\circ$). Our simulations are in agreement with $\alpha=1$. It is important to point out that ref.~\cite{Macedo2021} reported a change from $\alpha=1$ to $\alpha=0.5$ for Cs vapor as $n$ increases, due to Lorentzian wings in the Voigt profile from a certain value of the vapor density. Also, no collision effects were observed in the data of ref.~\cite{Baudouin2014}, for a Rb vapor in a thin disk-shaped cell with densities up to $n=1 \times 10^{20}$ atoms/m$^3$.

%---------------------------------------------
\subsection{Experimental results}
\label{subsec_results}
%\subsection{Temporal dynamics of the diffuse transmission and reflection}

% Fig 3a: I_{for}(t) vs t, several temperatures
% Fig 3b: tau_fit vs n, from I_for(t)
\begin{figure}[t]
\centering
\includegraphics[scale=1]{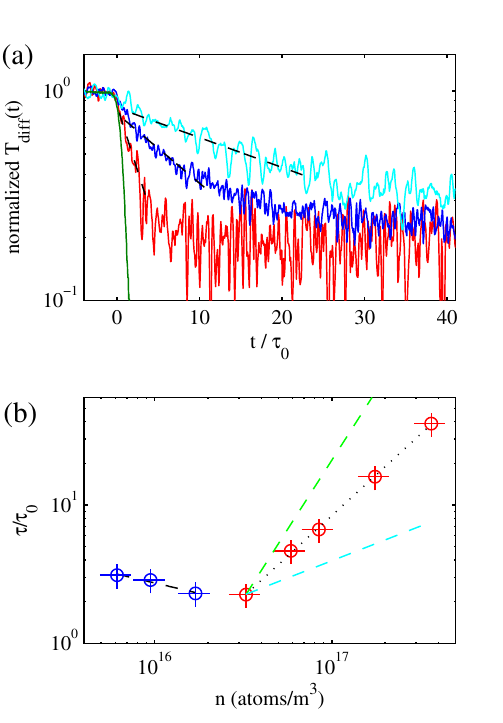}
\caption{(a) Normalized temporal diffuse transmission $T_\mathrm{diff}(t)$ as a function of the normalized time $t/\tau_0$ for three different temperatures: $30^\circ \text{C}$ ($n=9.5\times 10^{15}$ atoms/m$^3$; red), $70^\circ \text{C}$ ($n=1.8\times 10^{17}$ atoms/m$^3$; blue) and $80^\circ \text{C}$ ($n=3.7\times 10^{17}$ atoms/m$^3$; cyan). The curves were obtained with the setup of Fig.~\ref{fig1}(b). The dashed black lines are the respective theoretical fitting curves in the interval $I(t)/I_0\in[0.3, 0.8]$, and the green fast decay is the switch off of the laser. (b) Decay times $\tau/\tau_0$ as a function of the density $n$ from $T_\mathrm{diff}(t)$ shown in panel (a). The different colors separate low and high densities. The dashed black lines are the respective theoretical adjusts using Eq.~\ref{tau_fit}. The dashed green and cyan lines are, respectively, the scalings with normal diffusion ($\tau\sim n^2$) and collisional L\'evy flights ($\tau\sim n^{0.5}$).}
\label{fig3}
\end{figure}

% Fig 4b: I_{back}(t) vs t, several temperatures
%\begin{figure}[h]
%\centering
%\includegraphics[scale=0.7]{figure4.pdf}
%%\includegraphics[scale=1]{figure4a_eu.pdf}
%\caption{Normalized temporal diffuse reflection as a function of the time for three different temperatures: $20^\circ \text{C}$ (blue line), $80^\circ \text{C}$ (green) and $125^\circ \text{C}$ (yellow). The dashed lines are the theoretical fitting curves in the interval $I(t)/I_0\in[0.3, 0.8]$.}.
%\label{fig4b}
%\end{figure}

From the setup of Fig.~\ref{fig1}(b), we extract the temporal diffuse transmission $T_\mathrm{diff}(t)$  as a function of the time $t$, for a given density $n$. Typical curves are shown in Fig.~\ref{fig3}(a), along with the AOM switch-off profile without vapor cell. The curves were normalized by their mean value in the range $t/\tau_0\in[-12,-9]$ (i.e., before the AOM switch off) and correspond to $\approx 200$ points. The time $t$ is normalized by the Rb natural time decay rate $\tau_0=1/\Gamma\approx 26$ ns. At the lowest temperature, $T=20^\circ$C [density $n=4.6\times 10^{15}$ atoms/m$^3$; not shown in Fig.~\ref{fig2}(a)], density is not high enough and the decay is approximately equal to the laser switch off extinction generated by the AOM, which is $16$ ns. For all temperatures, we identify an exponential decay in the range $T_\mathrm{diff}(t)\in[0.3, 0.8]$. %, where $I(t)=\{ T_\mathrm{diff}(t), R_\mathrm{diff}(t) \}$.
%
%For $T_\mathrm{diff}(t)$, 
The data were acquired from $T=24^\circ$C ($n=6.1\times 10^{15}$ atoms/m$^3$) to $T=80^\circ$C ($n=3.6\times 10^{17}$ atoms/m$^3$). For higher temperatures, few photons are collected and we have a bad signal-to-noise ratio, and consequently no data in the mentioned range. %On the other hand, for $R_\mathrm{diff}(t)$, we have achieved temperatures of $T=145^\circ$C ($n=4.2\times 10^{19}$ atoms/m$^3$). Also, we checked that $R_\mathrm{diff}(t)$ curves saturate and collapse from $T > 115^\circ$C ($n > 4.7\times 10^{18}$ atoms/m$^3$) [not shown in Fig.~\ref{fig4}(a)].
Note that the total amount of forward diffuse photons, given by Eq.~\ref{Tdiff_vs_n} for a slab geometry~\cite{Baudouin2014}, tend to zero for large $n$.%, which means that the total amount of backward diffuse photons tend to the input photon number. %In previous works}, this is called the Ohm's law for photons~\cite{VanRossum1999}, studied in cold atoms~\cite{Labeyrie2004}, where we have normal diffusion~\cite{Labeyrie2003} and $\alpha=2$ in Eq.~\ref{Tdiff_vs_n}}.

%As discussed in section~\ref{sec_sim}, this is due to the increasing number of photons which are backward scattered, which tends to the total amount of input photons if $n\rightarrow \infty$. %This number increases with the atomic density (in fact optical depth) and tends to the total amount of the scattered photons if $n\rightarrow\infty$.

The temporal decay rate of these curves were extracted by using an exponential function $I(t) = A e^{-t/\tau}$~\cite{Baudouin2014, Savo2014, Klinger2023}, where $A$ and $\tau$ were kept as free parameters. Then, we plot $\tau$ \textit{versus} the density $n$. A theoretical fitting is done using a power-law equation~\cite{Klinger2023}

\begin{equation}
\tau = B n^\alpha\,,
\label{tau_fit}
\end{equation}
\\*
where $B$ and $\alpha$ were kept as free parameters.

The experimental data for $\tau$ and the corresponding fitting curves are given in Fig.~\ref{fig3}(b).% and~\ref{fig4}(b) for the diffuse transmission and reflection, respectively. 
We see two dominant regions: for low density, we have $\alpha \approx 0$, i.e., a constant region, whereas for high densities $\alpha$ is close to unity. For the low-density region, the resonant mean free path, defined by

\begin{equation}
\ell=\dfrac{1}{n\sigma_0f(0)}
\label{ell0}
\end{equation}
\\*
with $n$ the vapor density, $\sigma_0$ the resonant cross section and $f(0)$ the Eq.\ref{eq_fV} with $\Delta=0$) is in the order of the cell length: for the forward scattered photons [Fig.~\ref{fig3}(b)], we have $\ell=1.5$ cm for $n=2\times 10^{16}$ atoms/m$^3$, thus many photons leave the vapor after performed a single scattering event. For the high-density region, we extract $\alpha= 1.2\pm 0.2$ for the transmission data, which is in agreement with the expected $\alpha=1$ and confirms that there are signatures of L\'evy flights from the forward fluorescence. This is one of the results of this work. Note that these data are also in agreement with the frequency-dependent diffuse transmission (Fig.~\ref{fig2}). We also checked that our data is consistent with the predictions from radiative transport equation~\cite{Holstein1947}, namely, $\tau/\tau_0\propto n\sqrt{\ln n}\sim n$ for large $n$.

%\red{Concluir que vemos voos de Levy em ambos os casos}

%\red{Acho que tem que falar sobre o valor de $\tau_{max}=20$ a $50$ no experimento e isso corresponde a, digamos, $1\%$ dos fotons que ficaram esse tempo todo no vapor. Mas o $\tau$ medido \'e a contribuicao desses fotons, ainda que eles nao sejam tantos.}

%\red{VER: contrastar com mov. browniano}

It is interesting to compare our results with the temporal dynamics for normal diffusion, already studied for cold atoms~\cite{Labeyrie2003, Labeyrie2005}. In these works, a cloud of Rb vapor  at temperatures of $\sim 200$ $\mu$K is excited by laser pulses. Doppler effect is not enough to redistribute the photon frequency, and the constant decay $\tau$, also obtained from an exponential fit of the emitted fluorescences, scales with $\tau\sim b_0^2$, where $b_0$ is the optical depth of the medium. This is equivalent to $\tau\sim n^\alpha$ with $\alpha=2$. Fig.~\ref{fig3}(b) displays the normal diffusion scaling for comparison. We also display the scaling for a collisional-broadened vapor, which leads to $\alpha=0.5$~\cite{Araujo2021}.

Our values of $\tau$ also provide a measurement of a first-passage observable called number of scattering events, i.e., the typical number of steps that a photon performs before leaving the vapor. Note that the time the photon spent in the vapor is given by

\begin{equation}
\tau=N_{sc}\tau_0
\label{tau_and_Nsc}
\end{equation}
\\*
since the travel time between two scattering events is negligible. In another words, we obtained $\tau/\tau_0=N_{sc}\propto n^\alpha$ with $\alpha=1$ in a bounded system. The maximum value we reached is in the order of $\sim 40$, for our highest density [see Fig.~\ref{fig3}(b)]. For L\'evy glasses, this quantity was not determined in~\cite{Savo2014}, probably because the L\'evy walk character does not allow to achieve the duration of each scattering event.

%We mention that the temporal dynamics were studied for cold atoms [citation needed], from where $\tau$ was obtained for temporal dynamics and observed normal diffusion.

%---------------------------------------------
\section{Simulations}
\label{sec_sim}

The observation of L\'evy statistics from the forward fluorescence (Fig.~\ref{fig3}), discussed in the last section, is due to the L\'evy flights of photons and it is in agreement with the observed data from the steady frequency-dependent fluorescence (Fig.~\ref{fig2}). %At the same time, Eqs.~\ref{Tdiff_temp} and~\ref{tau_Ld} are derived for a one-dimensional L\'evy flight, and Eq.~\ref{Tdiff_vs_n} is derived for a infinite slab,} whereas our real geometry is a turned finite cylinder.}
At the same time, Eqs.~\ref{Tdiff_temp},~\ref{tau_Ld} and~\ref{Tdiff_vs_n} are derived for one-dimensional L\'evy flight, whereas our real geometry is a turned finite cylinder.

To address these issues, we perform Monte Carlo simulations to compare with our data. The simulations methods are discussed elsewhere~\cite{Carvalho2015}. In simplified words, we send a number $N_{ph}$ of photons with detuning $\Delta=0$ into a vapor with density $n$ and temperature $T$. For each photon, we draw the step length $x$ the photon walks before finding a two-level atom. The probability distribution of $x$ for a given detuning is $P_\Delta(x)=e^{-k(\Delta)x}$, where $k(\Delta)$ is the vapor absorption coefficient given by $k(\Delta)=n\sigma_0 f(\Delta)$ and Eq.~\ref{eq_fV}. After finding an atom, we draw randomly the scattering direction of the photon and calculate its new $\Delta$, where $\Delta$ is redistributed by Doppler effect and depends on the atom velocity components, which are drawn from a Maxwell-Boltzmann distribution with temperature $T$. Then, we draw $x$ from $P_\Delta(x)$ for this new detuning, and we repeat the steps until the photon leaves the cell. The cell is modelled as in the experiment, with same dimensions and orientation (see subsection~\ref{subsec_setup}). After the photon leaves the cell, we save its number of scattering events $N_{sc}$ and its output detuning $\Delta$.

The method described above does not take into account atomic collisions. To simulate collisional effects, we replace $\Gamma$ in Eq.~\ref{eq_fV} by $\Gamma_\mathrm{tot}=\Gamma+\Gamma_\mathrm{col}$, where $\Gamma_\mathrm{col}$ is the collisional rate, given by Eq.~\ref{gamma_c}. Then we define  the collision probability as $P_\mathrm{col}=\Gamma_\mathrm{col}/(\Gamma+\Gamma_\mathrm{col})$~\cite{Macedo2021}, and we draw a number between $0$ and $1$ to compare with $P_\mathrm{col}$. If a collision does not occur, we calculate the new detuning $\Delta$ as before. If a collision occurs, we draw the new photon detuning $\Delta$ from a Lorentzian distribution of width $\Gamma_\mathrm{col}$ and then we correct it by Doppler effect only in the emission.

We also save the emitted direction $\theta$ for each photon after it leaves the cell, where $0^\circ \leq\theta\leq 180^\circ$, measured from the incidence direction. If $\theta\in[0^\circ,90^\circ]$, the photon is diffuse transmitted. The diffuse transmission $T_\mathrm{diff}$ is thus proportional to the number of photons emitted in this range of $\theta$. %\blue{In the end of section \ref{subsec_setup}, we have plotted $T_\mathrm{diff}$ with $n$ and done a fit with Eq.~\ref{Tdiff_vs_n}. We obtained $\alpha=1$ for the mentioned collisional ranges.}

%\blue{
\subsection{Comparison between 1D and 3D geometries}
%}

The validity of Eqs.~\ref{Tdiff_temp},  \ref{tau_Ld} and \ref{Tdiff_vs_n} for tridimensional systems is not obvious, since these equations are derived for L\'evy flights in a infinite slab and/or one-dimensional geometry. Analytical theories for 3D L\'evy flights and geometry effects are in general scarce~\cite{Zaburdaev2015}.

\begin{figure}[h]
\centering
\includegraphics[scale=1]{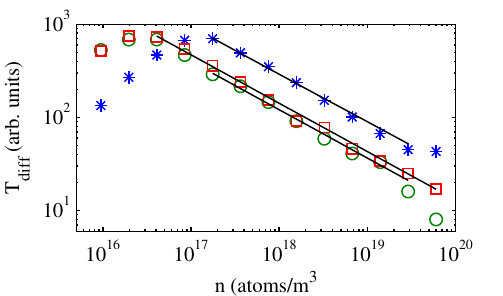}
\caption{Diffuse transmissions $T_\mathrm{diff}$ as a function of the vapor density $n$, for three different geometries: a thin cylindrical slab with radius 5 cm and thickness 5 mm (blue $\ast$), a thick cylindrical slab of radius 5 cm and thickness 2.5 cm (green $\circ$), and a turned cylinder with radius 1.25 cm and length 8 cm (red $\square$). The black full lines are fitting curves with Eq.~\ref{Tdiff_vs_n} in the range where a power law decay is observed. The simulations were done in the non-collisional regime.}
\label{fig8}
\end{figure}

For a direct comparison, we ran simulations for same vapor densities, but for three different geometries: (i) a finite thin slab geometry, modelled by a disk shaped cell with same dimensions as in ref.~\cite{Baudouin2014} (radius 5 cm and length 5 mm); a thick disk-shaped cell (radius 5 cm but length 2.5 cm); and (iii) our real geometry, a turned cylinder as described in section~\ref{subsec_setup}, with radius 1.25 cm and length 8 cm. We display the results in Fig.~\ref{fig8} in the non-collisional regime. The three fitting curves give, respectively, $\alpha=1.02\pm 0.06$, $\alpha=1.0\pm 0.1$ and $\alpha=1.04\pm 0.07$, confirming that the scaling of $T_\mathrm{diff}$ \textit{versus} $n$ [c.f. Fig.~\ref{fig2}(b)] is consistent with $\alpha=1$.

% Fig 8: T_diff vs n for several geometries

% Mean free path at the peak density:
% Thin slab: 1.8 mm (70ºC, n=1.8e17); 3.2 cm at 30ºC
% Thick slab: 7.6 mm (50°C, n=4e16); 3.2 cm at 30ºC
% Ricardo: 7.6 mm (50°C, n=4e16); 3.2 cm at 30ºC

For the three geometries, $T_\mathrm{diff}$ increases for small $n$ and then decreases. This behavior can be interpreted as the mean free path (defined in Eq.~\ref{ell0}) compared with the cell thickness. For the thick slab (thickness $L=2.5$ cm) and the turned cylinder (radius $R=1.25$ cm, which gives a thickness of $2R=2.5$ cm), the transition in $T_\mathrm{diff}$ occurs at $n_0=4.0\times 10^{16}$ atoms/m$^3$, which gives $\ell=7.6$ mm, i.e., $\ell/L=\ell/(2R)=0.3$. A similar behavior occurs with the thin slab (thickness $L=5$ mm): the peak in $T_\mathrm{diff}$ occurs at $n_0=4.0\times 10^{16}$ atoms/m$^3$, where $\ell=1.8$ mm and $\ell/L=0.36$. For $n<n_0$, we have few scattering events, while for $n>n_0$, many scattering events take place and the decrease of $T_\mathrm{diff}$ leads to signatures of L\'evy flights.

\subsection{Single and multiple scattering in the decay dynamics}

In order to simulate the temporal emitted fluorescences, we assume that the $N_{ph}$ photons enter at the cell at the same time, and thus we obtain a histogram of the number of exiting photons as a function of $N_{sc}$. This is because the photon spent a time $\tau$ in the vapor and it is given by Eq.~\ref{tau_and_Nsc}. To simulate the effect of a square pulse of duration $\Delta t_p$ driving the vapor, we make a convolution of the obtained histogram and a square pulse~\cite{Labeyrie2005}. Then, we normalize its maximum value to unity.

%\subsection{xxx}

%[To check the possibility of single scattering or even few scattering events before the photon leaves the cell dor the diffuse reflection, interpret the region at low density, ... .]

% Fig 5a: I_{for}(t) and I_{back}(t) from simulations
% Fig 5b: tau_fit vs n from simulations
\begin{figure}[h]
\centering
\includegraphics[scale=1]{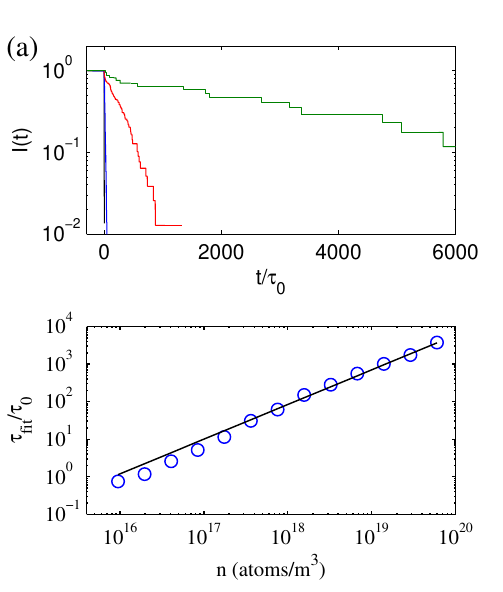}
\caption{(a) Normalized temporal  fluorescences as a function of the time for $30^\circ \text{C}$ (black; $n=9.5\times 10^{15}$ atoms/m$^3$), $70^\circ \text{C}$ (blue; $n=1.8\times 10^{17}$ atoms/m$^3$), $110^\circ \text{C}$ (red; $n=3.3\times 10^{18}$ atoms/m$^3$) and $150^\circ \text{C}$ (green; $n=6.0\times 10^{19}$ atoms/m$^3$). %The black full lines are the fitting curves.}
(b) Normalized decay times $\tau/\tau_0$ as a function of the density $n$ (blue circles). The full black line is a fitting curve with Eq.~\ref{tau_fit}.}
% theoretical curve $\tau/\tau_0\propto n^\alpha$ with $\alpha=1$.
\label{fig5}
\end{figure}

Figure~\ref{fig5}(a) compares directly the normalized temporal diffuse transmissions, for several temperatures. At low temperatures (i.e., low $n$), the curves have a single predominant exponential decay. At high temperatures (i.e., high $n$), this single exponential decay remains. We adjust these decays with an exponential law in each time range in order to extract the time decay rates.

The time decay rates are plotted in Fig.~\ref{fig5}(b). The fit range is $T_\mathrm{diff}(t)\in[0.01, 0.9]$. A fit with Eq.~\ref{tau_fit} leads to $\alpha=0.92\pm 0.05$ [black line in Fig.~\ref{fig5}(b)]. We check that the value of $\alpha$ does not vary when changing the fit interval to a ``early'' interval $T_\mathrm{diff}(t)\in[0.1, 0.9]$ (i.e., high amplitudes, corresponding to early times $t$ and to a ``late'' interval $T_\mathrm{diff}(t)\in[0.01, 0.1]$ (i.e., for lower amplitudes, corresponding to late times $t$). % For these intervals, we obtain, respec $\alpha=???\pm ???$ and $\alpha=???\pm ???$, respectively.} 
%\red{VER: For comparison, a curve $\tau\propto n^1$ is also shown.} 
We find total agreement with our experiment. Note that the simulations were performed with two-level atoms, and due to this we do not expect quantitative agreement.

% Fig 6: ratio Nsc/total(Nsc), I_{for}(t)
\begin{figure}[h]
\centering
\includegraphics[scale=1]{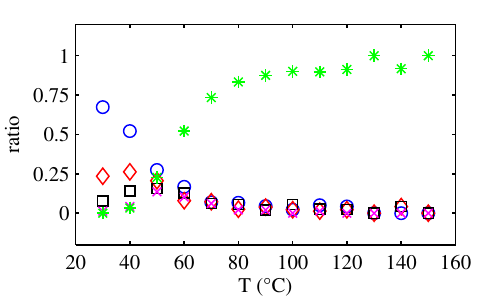}
\caption{Ratio between the amount of scattered photons (ratio S.S.) for a fixed $N_\mathrm{sc}$ and the total amount of scattered photons for a fixed temperature, for (a) diffuse transmission and (b) diffused reflection. Legend: $N_{sc}=1$ (blue $\circ$), $N_{sc}=1$ (red $\diamond$), $N_{sc}=3$ (black $\square$), $N_{sc}=4$ (magenta $\times$) and $N_{sc}\geq 5$ (green $\ast$).}
% symb={'ob','dr','sk','xm','+g'};
% $\circ$, $\diamond$, $+$, $\ast$, $\times$, $\square$
\label{fig6}
\end{figure}

\subsection{Role of single and multiple scattering in the temporal diffuse transmission}

At low density $n$, many photons leave the vapor with low scattering event number, typically $N_{sc}=1$. As $n$ increases, the typical resonant step length $\ell$ between two scattering events, given by Eq.~\ref{ell0}, decreases, so more photons perform more scattering events before leaving the vapor.

We quantify the number of scattering events performed by the photons for diffuse transmission after they left the vapor. Fig.~\ref{fig6} displays the results. The vertical axis is the number of photons which left the vapor with a given $N_{sc}$ divided by the total amount of photons scattered in the same direction, for a fixed temperature. Temperatures go from $30^\circ$C to $150^\circ$C, which implies in densities from $n=9.5\times 10^{15}$ to $n=6.0\times 10^{19}$ atoms/m$^3$. We see that the amount of photons which performed low scattering events ($N_{sc} \leq 4$) decreases quickly for increasing temperature $T$, becoming zero for $T \geq 100^\circ$C ($n \geq 1.6 \times 10^{18}$ atoms/m$^3$). This means that in this range almost all photons ($0.9$, i.e., $90\%$) leave the vapor performing $N_{sc} \geq 5$ scattering events. This means multiple scattering, complete frequency redistribution and consequently a single decay for $T_\mathrm{diff}(t)$, observed in both experimental and theoretical results. %Despite of the low scattering events for low density, this does not seem to impact the global behavior of the decay time $\tau$ (see Fig.~\ref{fig5}(b), where $\tau$ scales with density in order to give $\alpha=1$.} \red{VER: Note that the experimental time decay rates $\tau$ plotted in Fig.~\ref{fig3}(b) are indeed a measurement of $N_{sc}$: for the highest density, we have $\tau/\tau_0=N_{sc}=40$. In the corresponding theoretical values are in the order of $1,000$ (Fig.~\ref{fig5}).}

\section{Conclusions}

We have studied the temporal dynamics of L\'evy flights in annealed disorder, and we have shown that it is possible to see L\'evy signatures from the temporal forward emitted fluorescence (diffuse transmission). Correlations between the photon steps are absent and the photon random walk can be modelled as a L\'evy flight. We observed in this regime the L\'evy parameter $\alpha=1$, which is consistent with a measurement of this same diffusion transmission in the stationary regime. Monte Carlo simulations confirm our findings. All simulations were done for two-level atoms, as in previous works~\cite{Mercadier2013}, and the results are found to be consistent. The impact of the atomic multilevel structure was studied recently~\cite{Nunes2025}. %, but no significant impact was observed for previous data \red{ver essa frase}. \red{ver mais}

%For increasing density vapor, the photons tend to be scattered backward, which represents a significant improvement of signal-to-noise ratio compared to diffuse transmission. However, we have a significant amount of photons emitted after few scattering events, specially single scattering, in contrast with the forward fluorescence where almost all photons have performed multiple steps. It is surprising that a collected fluorescence is composed of half of photons performing a single scattering event and the few multiple scattered photons dominate the observed value of $\alpha$. %We attribute this %before leaving the vapor, where normally ] [but the amount of photons which performs multiple scattering dominate the observed statistics.] A short pulses

The temporal dynamics of L\'evy-like random walks was studied so far for quenched disorder, where L\'evy walks instead of L\'evy flights take place. Also, our simulations show that L\'evy flights can be observed for several finite three-dimensional geometries, like thin and thick cylinders and a turned long cylinder as reported in this work. Experimental signatures of L\'evy flights for vapor in thin and thick disks are already reported~\cite{Baudouin2014, Araujo2021, Macedo2021}. This represents a breakthrough in the state of art of the L\'evy flights as a whole, since analytical results in three dimensions are still scarce~\cite{Zaburdaev2015}, although with some recent theoretical works~\cite{Klinger2025, Chalas2026}.

This work is a first investigation of the L\'evy parameter in atomic media via temporal dynamics of the emitted fluorescence. Extensions of this work are twofold. On one hand, to include the short excitation regime, where the excitation duration is smaller than the atomic lifetime. In this context, normal diffusion has been observed in biological tissue~\cite{Chernomordik2002}, and L\'evy diffusion was observed in L\'evy glasses~\cite{Savo2014}. On another hand, to investigate if signatures of L\'evy flights are observed via backscattered light, where the number of photons increases and should improve the signal-to-noise ratio. We hope this work can contribute to further investigations of photon diffusion.

\section*{Acknowledgements}

We thank Sandra S. Vianna for borrowing the cell oven and detectors, and Naudson L. L. Matias and Nicolas A. Pessoa for helping on the experimental setup and data analysis, respectively. We also acknowledge the financial support from UFPB, UFPE and the agencies: Army Research Office (Grant No. W911NF-23-1-0287), Coordena\c{c}\~ao de Aperfei\c{c}oamento de Pessoal de N\'ivel Superior (CAPES), Conselho Nacional de
Desenvolvimento Cient\'ifico e Tecnol\'ogico (CNPq), Funda\c{c}\~ao de Amparo \`a Ci\^encia do Estado de Pernambuco (FACEPE), FAPESP (2021/06535-0), National Quantum Information Institute (INCT-IQ, No. 410 465469/2014-0)/CNPq, and the National Photonics Institute (INCT-INFo, No. 409174/2024-6)/CNPq.

%%%%%%%%%%%%%%%%%%%%%%%%%%%%%%%%%%%%%%%%%%%%%%%%%%%%

%\bibliography{refs_papers_levy}

\begin{thebibliography}{50}

%\bibitem{Araujo2016} M. O. Ara\'ujo, I. Kre\v{s}i\'c, R. Kaiser, and W. Guerin, Physical Review Letters \textbf{117}, 073002 (2016).

\bibitem{Viswanathan1996} G. M. Viswanathan, V. Afanasyev, S. V. Buldyrev, E. J.
Murphy, P. A. Prince, and H. E. Stanley, Nature \textbf{381},
413 (1996).

\bibitem{Edwards2007} A. M. Edwards, R. A. Phillips, N. W. Watkins, M. P.
Freeman, E. J. Murphy, V. Afanasyev, S. V. Buldyrev, M. G. E. da Luz, E. P. Raposo, H. E. Stanley, and G. M. Viswanathan, Nature
\textbf{449}, 1044 (2007).
%André M. Edwards ,Richard A. Phillips ,Nicholas W. Watkins ,Mervyn P. Freeman ,Eugene J. Murphy ,Vsevolod Afanasyev ,Sergey V. Buldyrev ,MGE da Luz ,EP Raposo ,H. Eugene Stanley eGandhimohan M. Viswanathan 

\bibitem{Li2022} J. Li, L. Weng, J. Xie, J. Amrit, and A. Ramiere, Physical Review E \textbf{105}, 064123 (2022).

\bibitem{Smith2023} T. B. Smith and D. B. Weissman, G3: Genes, Genomes, Genetics \textbf{13}, jkad023 (2023).
%https://academic.oup.com/g3journal/articlepdf/
%13/4/jkad023/56716347/jkad023.pdf.

\bibitem{Chernomordik2002} V. V. Chernomordik, D. W. Hattery, D. Grosenick,
H. Wabnitz, H. H. Rinneberg, K. T. Moesta, P. M.
Schlag, and A. H. Gandjbakhche, Journal of Biomedical
Optics \textbf{7}, 80 (2002).

\bibitem{Labeyrie2003} G. Labeyrie, E. Vaujour, C. A. M\"uller, D. Delande,
C. Miniatura, D. Wilkowski, and R. Kaiser, Physical Review Letters \textbf{91}, 223904 (2003).

\bibitem{Ivanov2024} V. V. Ivanov and J. M. Dlugach, Journal of Quantitative
Spectroscopy and Radiative Transfer \textbf{323}, 108999 (2024).

\bibitem{Barthelemy2008} P. Barthelemy, J. Bertolotti, and D. S. Wiersma, Nature
\textbf{453}, 495 (2008).

\bibitem{Mercadier2009} N. Mercadier, W. Guerin, M. Chevrollier, and R. Kaiser,
Nature Physics \textbf{5}, 602 (2009).

\bibitem{Holstein1947} T. Holstein, Physical Review \textbf{72}, 1212 (1947).

\bibitem{Holstein1951} T. Holstein, Physical Review \textbf{83}, 1159 (1951).

\bibitem{AlvesPereira2007} A. R. Alves-Pereira, E. J. Nunes-Pereira, J. M. G. Martinho,
and M. N. Berberan-Santos, The Journal of Chemical
Physics \textbf{126}, 154505 (2007).

\bibitem{Chevrollier2012} M. Chevrollier, Contemporary Physics \textbf{53}, 227 (2012).
%https://doi.org/10.1080/00107514.2012.684481.

\bibitem{Macedo2021} A. S. M. Macedo, J. P. Lopez, and T. Passerat de Silans,
Physical Review E \textbf{104}, 054143 (2021).

\bibitem{Baudouin2014} Q. Baudouin, R. Pierrat, A. Eloy, E. J. Nunes-Pereira,
P.-A. Cuniasse, N. Mercadier, and R. Kaiser, Physical Review E \textbf{90}, 052114 (2014).

\bibitem{Araujo2021} M. O. Ara\'ujo, T. P. de Silans, and R. Kaiser, Physical Review E \textbf{103}, L010101 (2021).

\bibitem{Lopez2023} J. P. Lopez, A. S. M. Macedo, M. O. Ara\'ujo, and
T. Passerat de Silans, Physical Review A \textbf{107}, 013501 (2023).

\bibitem{Colbert1990} T. Colbert and J. Huennekens, Physical Review A \textbf{41}, 6145
(1990).

\bibitem{Savo2014} R. Savo, M. Burresi, T. Svensson, K. Vynck, and D. S.
Wiersma, Physical Review A \textbf{90}, 023839 (2014).

\bibitem{Zaburdaev2015} V. Zaburdaev, S. Denisov, and J. Klafter, Rev. Mod.
Phys. \textbf{87}, 483 (2015).

\bibitem{Klinger2025} J. Klinger, O. B\'enichou, and R. Voituriez, arXiv preprint arXiv:2506.12581 (2025).

\bibitem{Chalas2026} K. Chalas, F. Diakonos, and A. Kapoyannis, arXiv
preprint arXiv:2605.14861 (2026).

\bibitem{Pereira2004} E. Pereira, J. M. G. Martinho, and M. N. Berberan-
Santos, Physical Review Letters \textbf{93}, 120201 (2004).

\bibitem{Klinger2023} J. Klinger, R. Voituriez, and O. B\'enichou, Physical Review E
\textbf{107}, 054109 (2023).

%\bibitem{VanRossum1999} M. C. W. van Rossum and T. M. Nieuwenhuizen, Review of Modern Physics \textbf{71}, 313 (1999).

%\bibitem{Guerin2017} W. Guerin, M. Rouabah, and R. Kaiser,
%Journal of Modern Optics \textbf{64}, 895 (2017).
%https://doi.org/10.1080/09500340.2016.1215564.

\bibitem{Mercadier2013} N. Mercadier, M. Chevrollier, W. Guerin, and R. Kaiser,
Physical Review A \textbf{87}, 063837 (2013).

\bibitem{Steck2001} D. A. Steck, Rubidium D Line Data, http://steck.us/alkalidata, 2025.

\bibitem{Weller2011} L. Weller, R. J. Bettles, P. Siddons, C. S. Adams, and
I. G. Hughes, Journal of Physics B: Atomic, Molecular
and Optical Physics \textbf{44}, 195006 (2011).

%\bibitem{Labeyrie2004} G. Labeyrie, D. Delande, C. M\"uller, C. Miniatura, and
R. Kaiser, Optics Communications \textbf{243}, 157 (2004).%, ultra Cold Atoms and Degenerate Quantum Gases.

\bibitem{Carvalho2015} J. C. de A. Carvalho, M. Ori\'a, M. Chevrollier, H. L. D. de S. Cavalcante, and T. Passerat de Silans, Physical Review A \textbf{91}, 053846 (2015).

\bibitem{Labeyrie2005} G. Labeyrie, R. Kaiser, and D. Delande, Applied Physics
B \textbf{81}, 1001 (2005).

%\bibitem{Chevrollier2010} M. Chevrollier, N. Mercadier, W. Guerin, and R. Kaiser,
The European Physical Journal D \textbf{58}, 161 (2010).

\bibitem{Nunes2025} I. C. Nunes, M. O. Ara\'ujo, J. P. Lopez, and T. Passerat de
Silans, Journal of Quantitative Spectroscopy and Radiative
Transfer \textbf{343}, 109481 (2025).


\end{thebibliography}
%\bibliographystyle{apsrev4-1}

\end{document}